\documentclass{article}

\usepackage[dcu]{harvard}
\usepackage{a4}
\usepackage{epsf}

\citationmode{abbr}
\citationstyle{dcu}
\makeindex

\newcommand{\eq}{\begin{equation}}
\newcommand{\en}{\end{equation}}
\newcommand{\eqa}{\begin{eqnarray}}
\newcommand{\ena}{\end{eqnarray}}

\title{Quantitative analysis of a Schaffer collateral model}

\author{
\bfseries Simon Schultz \\
\bfseries Stefano Panzeri \\
\bfseries Edmund Rolls \\
\it University of Oxford, Department of Experimental Psychology \\
\it South Parks Rd., Oxford OX1 3UD, U.K. \\
\bfseries Alessandro Treves \\
\it Programme in Neuroscience, SISSA, \\
\it via Beirut 2-4, 34013 Trieste, Italy \\
}

\date{To appear in R. Baddeley, P. Hancock and P. F{\"o}ldi{\'a}k (Eds.)
{\em Information Theory and the Brain}, Cambridge University Press,
Cambridge, U.K., 1998.}

\begin{document}

\maketitle

\section{Introduction}

Recent advances in techniques for the formal analysis of neural
networks \cite{Ami+87,Gar88,Tso+88,Tre90,Nad+93} have introduced the
possibility of detailed quantitative analyses of real brain
circuitry. This approach is particularly appropriate for regions such
as the hippocampus, which show distinct structure and for which the
microanatomy is relatively simple and well known.  \index{hippocampus}

The hippocampus, as archicortex, is thought to pre-date phylogenetically
the more complex neocortex, and certainly possesses a simplified version
of the six-layered neocortical stratification. It is not of interest
merely because of its simplicity, however: evidence from numerous
experimental paradigms and species points to a prominent role in the
formation of long-term memory, one of the core problems of cognitive
neuroscience \cite{Sco+57,Wei87,Gaf92,Coh+93,McN+87,Rol91}. Much useful
research in neurophysiology and neuropsychology has been directed
qualitatively, and even merely categorially, at understanding hippocampal
function. Awareness has dawned, however, that the analysis of {\em
quantitative} aspects of hippocampal organisation is essential to an
understanding of why evolutionary pressures have resulted in the mammalian
hippocampal system being the way it is
\cite{Ama+90,Tre+96,Ste83,Wit+92}. Such an understanding will require a
theoretical framework (or formalism) that is sufficiently powerful to
yield quantitative expressions for meaningful parameters, that can be
considered valid for the real hippocampus, is parsimonious with known
physiology, and is simple enough to avoid being swamped by details that
might obscure phenomena of real interest.\index{archicortex}
\index{quantitative analysis}\index{formalism}

\begin{figure}
\epsfxsize=11cm
\epsfysize=18cm
\epsfbox{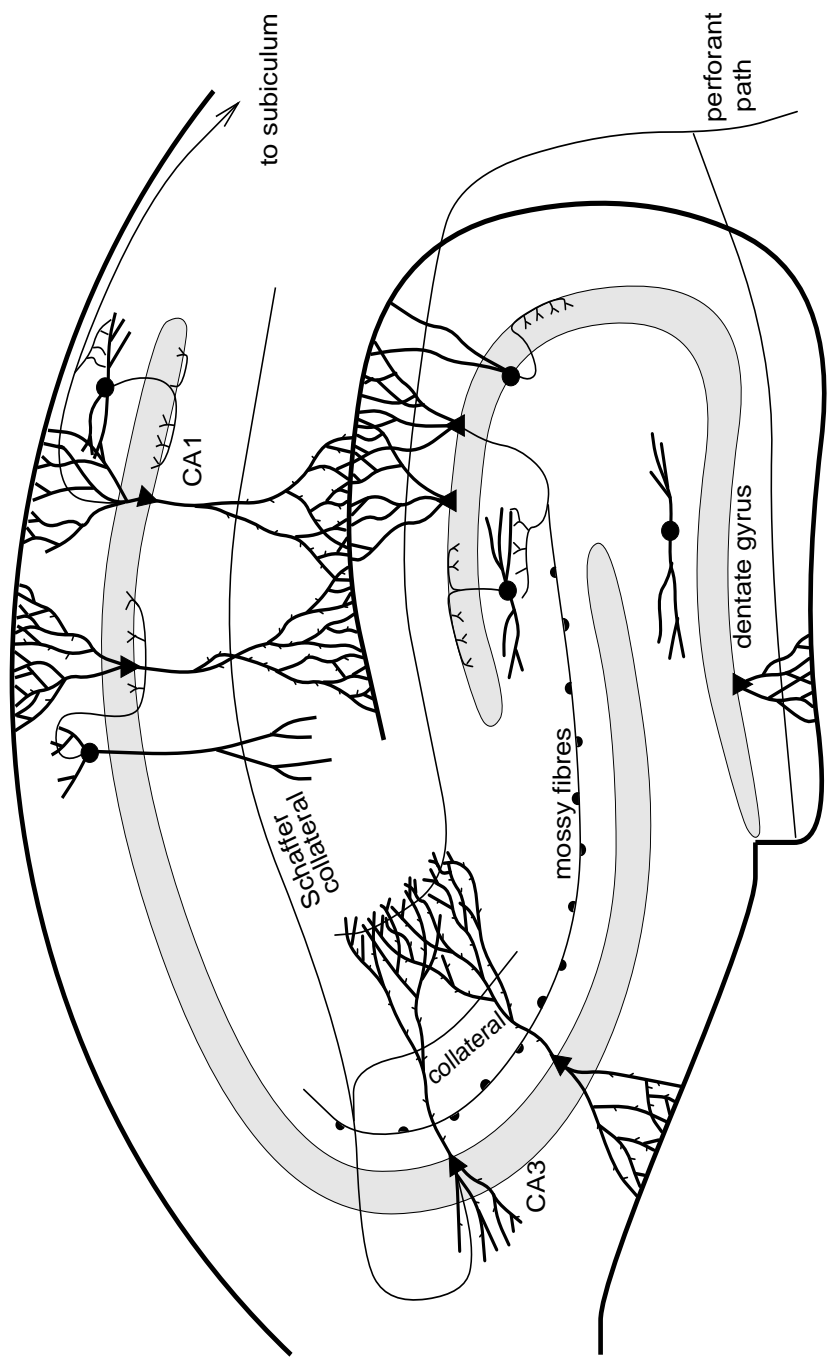}
\caption{A schematic diagram of the hippocampal formation. Information
enters the hippocampus from layer 2 entorhinal cells by the perforant
path, which projects into dentate gyrus, CA3 and CA1 areas. In addition to
its perforant path inputs, CA3 receives a lesser number of mossy fibre
synapses from the dentate granule cells. The axons of the CA3 pyramidal
cells project commissurally, recurrently within CA3, and also forward to
area CA1 by the Schaffer collateral pathway. Information leaves the
hippocampus via backprojections to the entorhinal cortex from CA1 and the
subiculum, and also via the fornix to the mammillary bodies and anterior
nucleus of the thalamus.}
\label{fig:sch-pic}
\end{figure}

The foundations of at least one such formalism were laid with the
notion that the recurrent collateral connections of subregion CA3 of
the hippocampus allow it to function as an autoassociative memory
(\citename{Rol89}, 1989, although many of the ideas go back to
\citename{Mar71}, 1971), and with subsequent quantitative analysis
(reviewed in \citename{Tre+94}, 1994). After the laying of
foundations, it is important to begin erecting a structural
framework. In this context, this refers to the modelling of further
features of the hippocampal system, in a parsimonious and simplistic
way. \citeasnoun{Tre95} introduced a model of the Schaffer
collaterals, the axonal projections which reach from the CA3 pyramidal
cells into subregion CA1, forming a major part of the output from CA3
and of the input to CA1. The Schaffer collaterals can be seen clearly
in Figure~\ref{fig:sch-pic}, a schematic drawing of the hippocampal
formation. This paper introduced an information theoretic
formalism similar to that of \citeasnoun{Nad+93} to the analysis. As
will become apparent, this approach to network analysis appears to be
particularly powerful, and is certain to find diverse application in
the future. \index{CA3 hippocampal subfield} \index{CA1 hippocampal
subfield} \index{Schaffer collaterals} \index{recurrent collaterals}

Once the rudiments of a structural framework have been erected, it is
possible to begin to add to the fabric of the theory -- to begin to
consider the effect of additional details of biology that were not in
themselves necessary to its structural basis. This is where the
contribution of the work described in this chapter lies. The analysis
described in \cite{Tre95} assumed, for the purposes of simplicity of
analysis, that the distribution of patterns of firing of CA3 pyramidal
neurons was binary (and for one case ternary), although it considered
threshold-linear (and thus analogue) model neurons. Here we shall consider
in more detail the effect on information transmission of the possible
graded nature of neuronal signalling. Another simple assumption made was
that the pattern of convergence (the number of connections each CA1 neuron
receives from CA3 neurons) of the Schaffer collaterals was either uniform,
or alternatively bi-layered. The real situation is slightly more complex,
and a better approximation of it is considered here.

\section{A model of the Schaffer collaterals}

The Schaffer collateral model describes, in a simplified form, the
connections from the $N$ CA3 pyramidal cells to the $M$ CA1 pyramidal
cells. Most Schaffer collateral axons project into the stratum radiatum of
CA1, although CA3 neurons proximal to CA1 tend to project into the stratum
oriens \cite{Ish+90}; in the model these are assumed to have the same
effect on the recipient pyramidal cells. Inhibitory interneurons are
considered to act only as regulators of pyramidal cell activity. The
perforant path synapses to CA1 cells are at this stage ignored (although
they have been considered elsewhere; see \citename{Ful+96}, this volume),
as are the few CA1 recurrent collaterals. The system is considered for the
purpose of analysis to operate in two distinct modes: storage and
retrieval. During storage the Schaffer collateral synaptic efficacies are
modified using a Hebbian rule reflecting the conjunction of pre- and
post-synaptic activity. This modification has a slower time-constant than
that governing neuronal activity, and thus does not affect the current CA1
output. During retrieval the Schaffer collaterals relay a pattern of
neural firing with synaptic efficacies which reflect all previous storage
events. In the following, the superscript $S$ is used to indicate the
storage phase, and $R$ to indicate the retrieval phase. \index{Schaffer
collaterals}

\begin{itemize}
\item $\{\eta_i\}$ are the firing rates of each cell $i$ of
CA3. The probability density of finding a given firing
pattern is taken to be:
\begin{equation}
P(\{\eta_i\}) = \prod_i P_{\eta}(\eta_i) d\eta_i
\end{equation} 
where $\eta$ is the vector of the above firing rates.
This assumption means that each cell in CA3 is taken to code for independent
information, an idealised version of the idea that by this stage most
of the redundancy present in earlier representations has been removed.
\item $\{V_i\}$ are the firing rates in the pattern retrieved from CA3, 
and they are taken to reproduce the $\{\eta_i\}$ with some Gaussian distortion 
(noise), followed by rectification
\begin{eqnarray}
V_i & = & [\eta_i +\delta_i]^+ \nonumber \\
\left<(\delta_i)^2\right> & = & \sigma^2_{\delta} 
\end{eqnarray}
(the rectifying function $[x]^+=x$ for $x>0$, and 0 otherwise, ensures that 
a firing rate is a positive quantity. This results in the probability
density of $V_i$ having a point component at zero equal to the sub-zero
contribution, in addition to the smooth component). $\sigma_{\delta}$ can
be related (e.g.) to interference effects due to the loading of other memory
patterns in CA3 (see below and \citename{Tre+91} 1991). This and the
following noise terms are all taken to have zero means.
\item $\{\xi_j\}$ are the firing rates produced in each cell $j$ of CA1, 
{\em during the storage} of the CA3 representation; they are determined by the
matrix multiplication of the pattern $\{\eta_i\}$ with the synaptic weights
$J_{ij}$ -- of zero mean, as explained below,  and variance $\sigma_J^2$ -- 
followed by Gaussian distortion, (inhibition-dependent) thresholding and rectification
\begin{eqnarray}
\xi_j&=&\left[ \xi_0 +\sum_i c_{ij} J^S_{ij} \eta_i +\epsilon^S_j
\right]^+\nonumber\\
\left<(\epsilon^S_j)^2\right> & = & \sigma^2_{\epsilon^S} \nonumber \\
\left<(J^S_{ij})^2\right> & = & \sigma^2_J.
\end{eqnarray}
The synaptic matrix is very sparse as each CA1 cell receives inputs from only
$C_j$ (of the order of $10^4$) cells in CA3. The average of $C_j$ across cells is with
each other.
denoted as $C$
\begin{eqnarray}
c_{ij}&=&\{0,1\} \nonumber \\
\left< c_{ij} \right> N &=& C_j \qquad (C\equiv \left<C_j\right>)
\end{eqnarray}
\item $\{ U_j\}$ are the firing rates produced in CA1 by the pattern 
$\{ V_i\}$ retrieved in CA3
\begin{eqnarray}
U_j & = & \left[ U_0 +\sum_i c_{ij} J^R_{ij} V_i +\epsilon^R_j \right]^+
\nonumber \\
\left<(\epsilon^R_j)^2\right>&=&\sigma^2_{\epsilon^R} \nonumber \\
\left<(J^R_{ij})^2\right>&=&\sigma^2_J
\end{eqnarray} 
Each weight of the synaptic matrix during retrieval of a specific pattern,
\begin{equation}
J^R_{ij}=\cos (\theta_{\mu})  J^S_{ij}+\gamma^{1/2}(\theta_{\mu}) 
H(\eta_i,\xi_j) +\sin (\theta_{\mu}) J^N_{ij}
\label{eq:fmod}
\end{equation}
consists of 
\begin{enumerate}
\item the original weight during storage, $J^S_{ij}$, damped by a factor
$cos(\theta_{\mu})$, where $0 < \theta_{\mu} < \pi / 2$
parameterises the time elapsed between the storage and retrieval of pattern
$\mu$ ($\mu$ is a shorthand for the pattern quadruplet
$\{\eta_i,V_i,\xi_j,U_j\}$ ).
\item the modification due to the storage of $\mu$ itself,
represented by a Hebbian term $H(\eta_i,\xi_j)$, normalised so that
\begin{equation}
\left<(H(\eta,\xi))^2\right>=\sigma^2_J;
\label{eq:e6}
\end{equation}
$\gamma$ measures the degree of {\em plasticity}, i.e. the mean square 
contribution of the modification induced by one pattern, over the overall 
variance, across time, of the synaptic weight.
\item the superimposed modifications $J^N$ reflecting the successive storage
of new intervening patterns, again normalised such that
\begin{equation}
\left<(J^N_{ij})^2\right>=\sigma^2_J.
\end{equation}
\end{enumerate}
\end{itemize}
The mean value across all patterns of each synaptic weight is taken to be
equal across synapses, and is therefore taken into the threshold term. The
synaptic weights $J_{ij}^R$ and $J_{ij}^S$ are thus of zero mean, and
variance $\sigma_J^2$ (all that affects the calculation is the first two
moments of their distribution). 

A plasticity model is used which corresponds to gradual decay of memory
traces. Numbering memory patterns from $1,...,\lambda,...,\infty$
backwards, the model sets $\cos(\theta_\lambda)$ = $exp(-\lambda\gamma_0/2)$
and $\gamma(\theta_\lambda) = \gamma_0 exp(-\lambda \gamma_0)$. Thus the
strength of older memories fades exponentially with the number of
intervening memories. The same forgetting model is assumed to apply to the
CA3 network, and for this network, the maximum number of patterns can be
stored when the plasticity $\gamma_0^{CA3} = 2/C$ \cite{Tre95}.

For the Hebbian term the specific form
\begin{equation}
H(\eta_i,\xi_j) = {h\over \sqrt{C}}(\xi_j-\xi_0)(\eta_i-\left<\eta_i\right>)
\end{equation}
is used, where $h$ ensures the normalisation given in Eq.~\ref{eq:e6}.
 
The thresholds $\xi_0$ and $U_0$ are assumed to be of fixed value in the
following analysis. This need not be the case, however, and as far as the
model represents (in a simplified fashion) the real hippocampus, they
might be considered to be tuned to constrain the sparseness of activity in
CA1 in the storage and retrieval phases of operation respectively,
reflecting inhibitory control of neural activity.

\begin{figure}
\epsfxsize=12cm
\epsfysize=8cm
\epsfbox{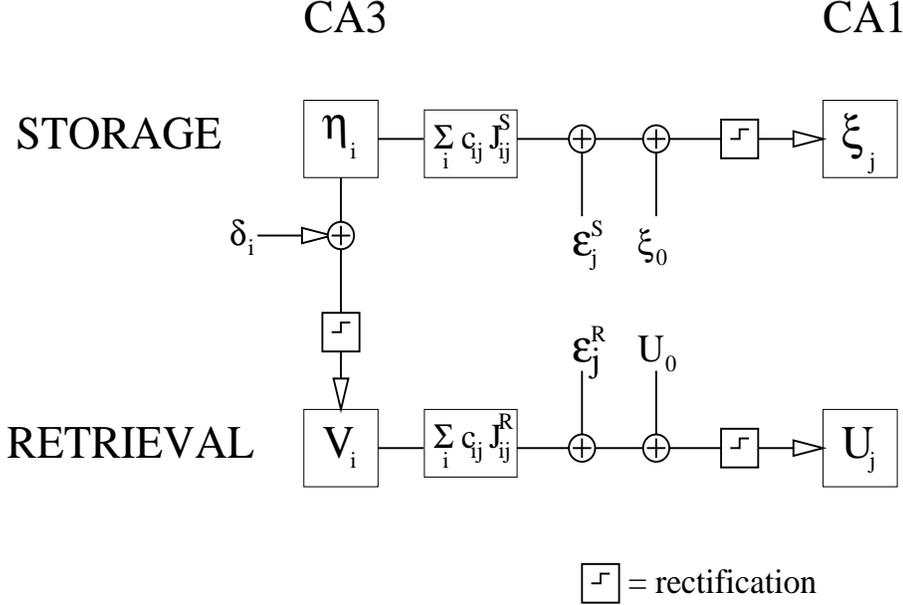}
\caption{A block diagram illustrating the relationships between the
variables present in the model. The input of the system could be
considered to be the CA3 pattern during storage, $\eta$, and the output
the CA1 pattern during retrieval, $U$. $J_{ij}^R$ depends on $J_{ij}^S$,
$\xi_j$ and $\eta_i$ as described in the text. }
\label{fig:block}
\end{figure}

The block diagram shown in Fig.~\ref{fig:block} illustrates the
relationships between the variables described in the preceding section.

\section{Technical comments}

The aim of the analysis is to calculate how much, on average, of the
information present in the original pattern $\{\eta_i\}$ is still
present in the effective output of the system, the pattern $\{ U_j\}$,
i.e. to average the mutual information
\begin{equation}
i(\{\eta_i\},\{ U_j\})=\int \prod_i d\eta_i \int \prod_j dU_j P(\{\eta_i\},
\{ U_j\}) \ln {P(\{\eta_i\},\{ U_j\})\over P(\{\eta_i\})P(\{ U_j\})}
\end{equation}
over the variables $c_{ij}, J^S_{ij}, J^N_{ij}$. The details of the
calculation are unfortunately too extensive to present here, and the
reader will have to be satisfied with an outline of the technique
used. Those not familiar with replica calculations may refer to the
final chapter of \cite{Her+91}, the appendices of \cite{Rol+97b}, or,
less accessibly, the book by \citeasnoun{Mez+87} for background
material.

$P(\{\eta_i\},\{ U_j\})$ is written (simplifying the notation) as
\begin{equation}
P(\eta, U ) = P(U \mid \eta) P(\eta ) = \int_V \int_{\xi} dV d\xi 
P(U\mid V, \xi, \eta) P(V\mid \eta) P(\xi \mid \eta) P(\eta)
\end{equation}
where the probability densities implement the model defined above.

The average mutual information is evaluated using the replica trick,
which amounts to
\eq
\log P = \lim_{n\rightarrow 0}\frac{(P^n-1)}{n}
\en
which involves a number of subtleties, for which \cite{Mez+87} can be
consulted for a complete discussion. The important thing to note is
that an assumption regarding replica symmetry is necessitated, and the
stability of resulting solutions must be checked. This has been reported
in \cite{Sch+97b} for a simplified version of the neural network analysed
here: a single layer of threshold-linear neurons with a
single phase of operation (transmission) rather than the dual (storage and
retrieval) modes in the model presented here. The conclusions of that
study were that the replica-symmetric solution is stable for sufficiently
sparse codes, but that for more distributed codes the solution was
unstable below a critical noise variance. These conclusions can be assumed
to carry across to the current model in at least a qualitative sense. In
those regions (low noise, distributed codes) where the replica-symmetric
solution is unstable, a solution with broken replica symmetry is
required. It should be noted that it is not known what quantitative
difference such a solution would bring: it may be very little, as is the
case with the Hopfield network\cite{Ami+87}.

The expression for mutual information thus becomes
\begin{equation}
\left< i(\eta,U) \right>_{c,J^S,J^N}=\lim_{n\to 0} {1\over n} \left< \int  d\eta dU P(\eta,U)
\left\{ \left[ P(\eta,U) \over P(\eta) \right]^n - \left[ P(U)\right]^n
\right\} \right>_{c,J^S,J^N}.\label{eq9}
\end{equation}
where it is necessary to introduce $n+1$ replicas of the variables $\delta_i,
\epsilon^S_j,\epsilon^R_j,V_i,\xi_j$ and, for the second term in curly
brackets only, $\eta_i$. 

The core of the calculation then is the calculation of the probability
density $\left< P(\eta,U)^{n+1}\right>$. \index{replica trick} The key to
this is ``self-consistent statistics'' (\citename{Rol+97b}, 1997, appendix
4): all possible values of each firing rate in the system are integrated,
subject to a set of constraints that implement the model. The constraints
are implemented using the integral form of the Dirac
$\delta$-function. Another set of Lagrange multipliers introduces
macroscopic parameters \eqa x^\alpha & = &
\frac{1}{N}\sum_i\frac{(\eta_i^\alpha-\left<\eta\right>)}
{\left<\eta\right>}V_i^\alpha\theta(V_i^\alpha) \nonumber\\
w^{\alpha\beta} & = & \frac{1}{N}\sum_i\eta_i^\alpha
V_i^\beta\theta(V_i^\beta) \nonumber\\ y^{\alpha\beta} & = &
\frac{1}{N}\sum_i V_i^\alpha V_i^\beta \theta(V_i^\alpha)
\theta(V_i^\beta) \nonumber\\ z^{\alpha\beta} & = &
\frac{1}{N}\sum_i\eta_i^\alpha\eta_i^\beta \nonumber\\ \ena where
$\theta(x)$ is the Heaviside function, and $\alpha,\beta$ are replica
indices. Making the assumption of replica symmetry, and performing the
integrals over all microscopic parameters, with some algebra an integral
expression is obtained for the average mutual information per CA3
cell. This integral over the macroscopic parameters and their respective
Lagrange multipliers is evaluated using the saddle-point approximation,
which is exact in the limit of an infinite number of neurons (see, for
example, \citename{Jef+72}, 1972) to yield the expression given in
Appendix A; the saddle-points of the expression must in general be found
numerically.

\section{How graded is information representation on the Schaffer collaterals?}

Specification of the probability density $P(\eta)$ allows different
distributions of firing rates in CA3 to be considered in the
analysis. Clearly the distribution of firing rates that should be
considered in the analysis is that of the firing of CA3 pyramidal cells,
computed over the time-constant of storage (which we can assume to be the
time-constant of LTP), during only those periods where biophysical
conditions are appropriate for learning to occur. Unfortunately this last
caveat makes a simple fit of the firing-rate distribution from single-unit
recordings fairly meaningless unless the correct assumptions regarding
exactly what these conditions are {\em in-vivo} can be made. It would be
fair to assume that cholinergic modulatory activity is a pre-requisite,
and unfortunately we cannot know directly from single-unit recordings from
the hippocampus when the cells recorded from are receiving significant
cholinergic modulation. Note that it might be possible to discover this
indirectly. In any event, possibly the most useful thing we can do for the
present is to assume that the distribution of firing rates during storage
is graded, sparse, and exponentially tailed. This accords with the
observations of neurophysiologists \cite{Bar+90,Rol+97b}. The easiest way
to introduce this to the current investigation is by means of a discrete
approximation to the exponential distribution, with extra weight given to
low firing rates. This allows quantitative investigation of the effects of
analogue resolution on the information transmission capabilities of the
Schaffer collateral model.

\begin{figure}
\begin{center}
\leavevmode
\epsfxsize=12cm
\epsfysize=6cm
\leavevmode
{\bfseries\Large a\epsfbox{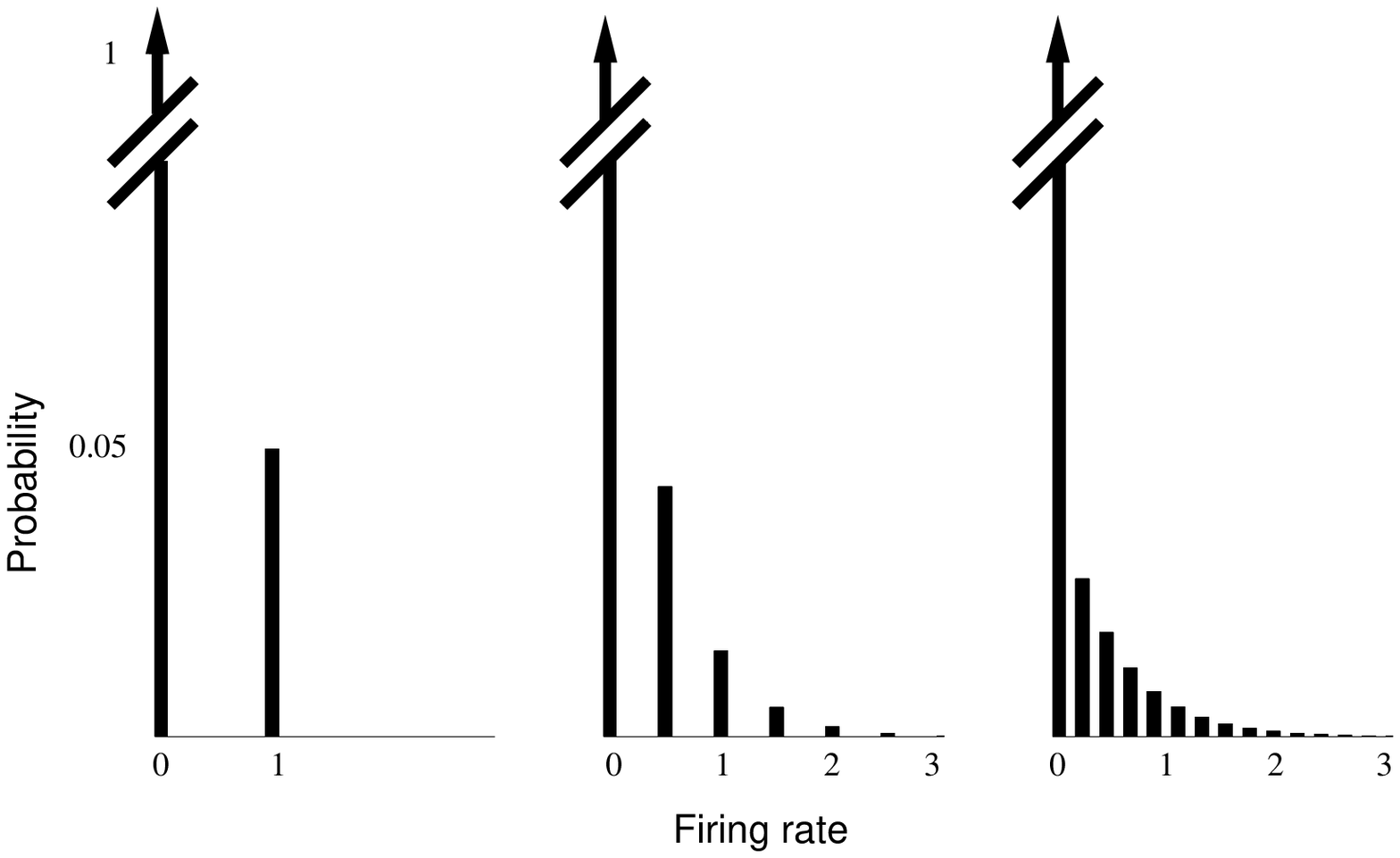}}
\end{center}
\begin{center}
\epsfxsize=6cm
\epsfysize=5cm
\leavevmode
\parbox[t]{6cm}{\bfseries\Large b\epsffile{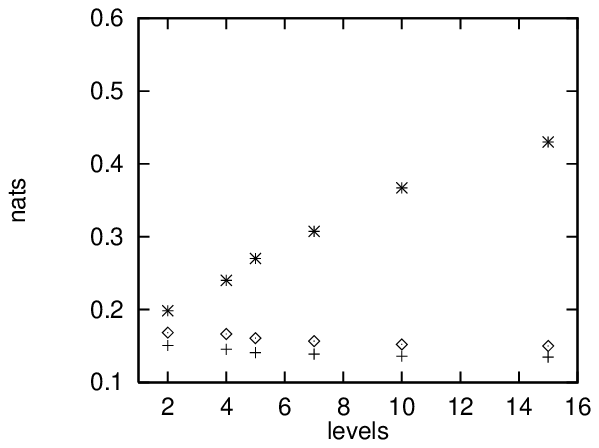}}
\epsfxsize=6cm
\epsfysize=5cm
\parbox[t]{6cm}{\bfseries\Large c\epsffile{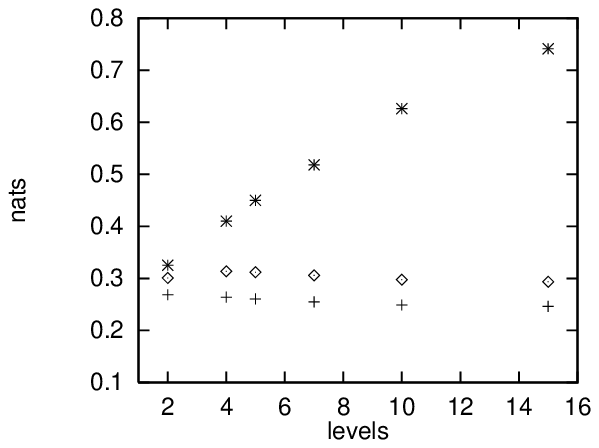}}
\epsfxsize=6cm
\epsfysize=5cm
\parbox[t]{6cm}{\bfseries\Large d\epsffile{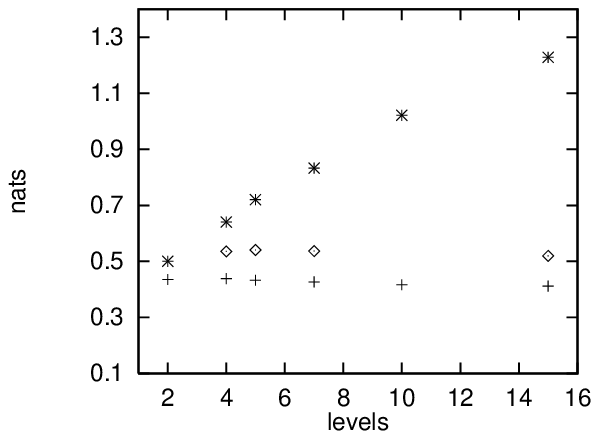}}
\epsfxsize=6cm
\epsfysize=5cm
\parbox[t]{6cm}{\bfseries\Large e\epsffile{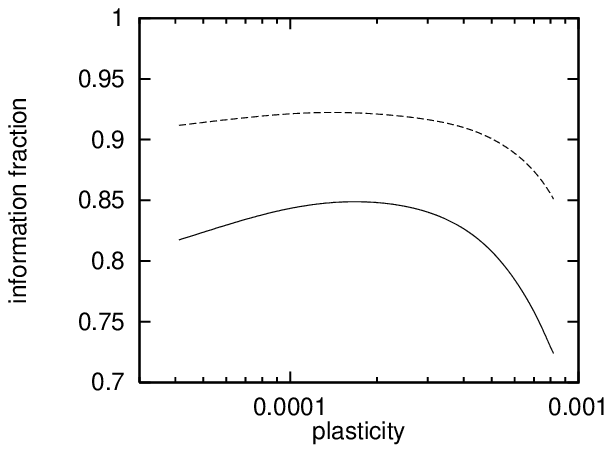}}
\caption{ {\bfseries a} Some of the CA3 firing rate distributions used in
the analysis. These are, in general, formed by the mixture of a unitary
distribution and a discretised exponential.  {\bfseries b} -- {\bfseries d}
The mutual information between patterns of firing in CA1 and patterns of
firing in CA3, expressed in natural units (nats). Asterisks represent the
entropy of the CA3 pattern distribution, diamonds the CA1 retrieved mutual
information, and crosses the CA1 information during the storage phase. The
horizontal axis parameterises the number of discrete levels in the input
distribution: for codes with fine analogue resolution, this is
greater. {\bfseries b} is for $a=0.05$ (sparse), {\bfseries c} for
$a=0.10$, and {\bfseries d} for $a=0.20$ (slightly more distributed).
{\bfseries e} The dependence of information transmission on the degree of
plasticity in the Schaffer collaterals, for $a=0.05$ (solid) and $a=0.10$
(dashed). A binary pattern distribution was used in this case.}
\label{fig:qlox}
\end{center}
\end{figure}

The required CA3 firing rate distributions were formed by the mixture
of the unitary distribution and the discretised exponential, using as
mixture parameters the offset $\epsilon$ between their origins, and
relative weightings. The distributions were constrained to have first
and second moments $\left<\eta\right>$, $\left<\eta^2\right>,$ and
thus sparseness $\left<\eta\right>^2/\left<\eta^2\right>$, equal to
$a$. In the cases considered here, $a$ was allowed values of 0.05, 0.10
and 0.20 only. The width of the distribution examined was set to 3.0,
and the number of discretised firing levels contained in this width
parameterised as $l$. The binary distribution was completely specified
by this; for distributions with a large number of levels, there was
some degree of freedom, but its numerical effect on the resulting
distributions was essentially negligible. Those distributions with a
small number of levels $\ge 2$ were non-unique, and were chosen fairly
arbitrarily for the following results, as those that had entropies
interpolating between the binary and large $l$ situations.  Some
examples of the distributions used are shown in Fig.~\ref{fig:qlox}a.

The total entropy per cell of the CA3 firing pattern, given a
probability distribution characterised by $L$ levels, is

\eq
\label{eq:maxinf}
h(\eta) = - \sum_{l=1}^L P_{\eta_l}(\eta_l) \ln P_{\eta_l}(\eta_l).
\en

The results are shown in Fig.~\ref{fig:qlox}b--d. The entropy present in
the CA3 firing rate distributions is marked by asterisks. The mutual
information conveyed by the retrieved pattern of CA1 firing rates, which
must be strictly less than the CA3 entropy, is represented by circles. It
is apparent that maximum information efficiency occurs in the binary
limit. More remarkably, even in absolute terms the information conveyed is
maximal for low resolution codes, at least for quite sparse codes. The
results are qualitatively consistent over sparsenesses $a$ ranging from
0.05 to 0.2; obviously with higher $a$ (more distributed codes), entropies
are greater. For more distributed codes (i.e. with signalling more evenly
distributed over neuronal firing rates), it appears that there may be some
small absolute increase in information with the use of analogue signalling
levels.

For comparison, the crosses in the figures show the information stored
in CA1. This was computed using a simpler version of the calculation,
in which the mutual information $i(\{\eta_i\},\{\xi_j\})$ was
calculated. Obviously, in this calculation, the CA3 and CA1 retrieval
noises $\sigma_\delta$ and $\sigma_\epsilon^R$ are not present; on the
other hand, neither is the Schaffer collateral memory term. Since the
retrieved CA1 information is in every case higher than that stored, we
can conclude that for the parameters considered, the additional Schaffer
memory effect outweighs the deleterious effects of the retrieval noise
distributions.

It follows from the forgetting model defined by Eq.~\ref{eq:fmod},
that information transmission is maximal when the plasticity (mean
square contribution of the modification induced by one pattern) is
matched in the CA3 recurrent collaterals and the Schaffer collaterals
\cite{Tre95}. It can be seen in Fig.~\ref{fig:qlox}e that this effect
is robust to the use of more distributed patterns. 

\section{Non-uniform Convergence}

It is assumed in \cite{Tre95} that there is uniform convergence of
connections from CA3 to CA1 across the extent of the CA1 subfield. In
reality, each CA1 pyramidal neuron does not receive the same number of
connections from CA3: this quantity varies across the transverse extent of
CA1 (although this transverse variance may be less than that within CA3;
\citename{Ama+90} 1990). \citename{Ber+94} (1994) investigated this with a
connectivity model constructed by simulating a {\em Phaseolus vulgaris}
leucoagglutinin labelling experiment, matched to the available anatomical
data. Their conclusion was that mid CA1 neurons receive more connections
(8000) than those in proximal and distal CA1 (6500). The precise numbers
are not important here; what {\em is} of interest is to consider the
effect on information transmission of this spread in the convergence
parameter $C_j$ about its mean $C$. 

In this analysis $\sigma_J^2$ is set to $1/C$ for all cells in the
network. $C$ is set using the assumption of parabolic dependence of $C_j$
upon transverse extent, on the basis of Fig.~5 of \cite{Ber+94}. In order
to facilitate comparison with the results reported in \citeasnoun{Tre95},
$C$ is held at 10,000 for all results in this section. The model used
(which we will refer to as the `realistic convergence' model) is thus
simply a scaled version of that due to \citename{Ber+94}, with $C_j =
7,143$ at the proximal and distal edges of CA1, and $C_j = 11,429$ at the
midpoint. Note that this refers to the number of CA3 cells contacting each CA1
cell; each may do so via more than one synapse.

\begin{figure}
\begin{center}
\epsfxsize=6cm
\epsfysize=5cm
\leavevmode
\parbox[t]{6cm}{\bfseries\Large a\epsffile{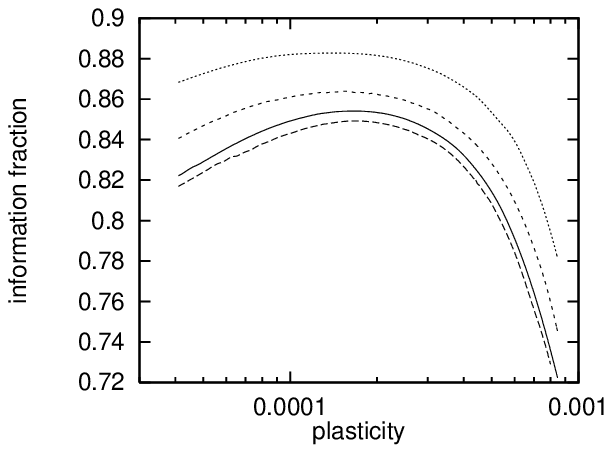}}
\epsfxsize=6cm
\epsfysize=5cm
\parbox[t]{6cm}{\bfseries\Large b\epsffile{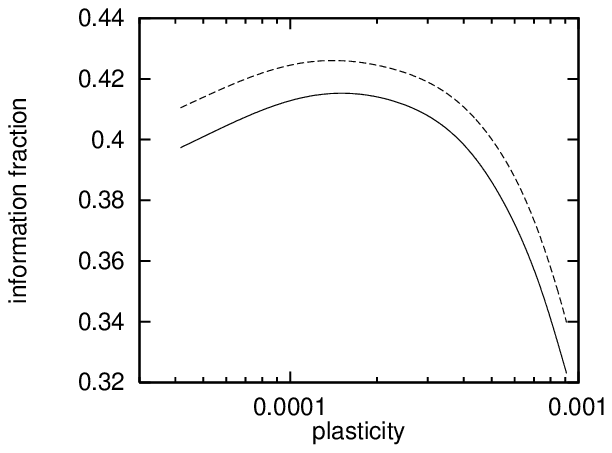}}
\end{center}
\caption{Information transmitted as a function of Schaffer collateral
plasticity. {\bfseries a} Binary CA3 firing rate distributions. The
solid line indicates the result for the realistic convergence
model. The dashed lines indicate, in ascending order: (i) uniform
convergence, (ii) two-tier convergence model with
$C_j\in\{5000,15000\}$, (iii) two-tier convergence model with
$C_j\in\{2000,18000\}$. {\bfseries b} With more realistic CA3
firing-rate distributions (the 10-level discrete exponential
approximation from the previous section). The solid line indicates the
result for uniform connectivity, and the dashed line the two-tier
convergence model with $C_j\in\{5000,15000\}$.}
\label{fig:plas}
\end{figure}

The saddle-point expression (\ref{eq:inf}) was evaluated numerically while
varying the plasticity of the Schaffer connections, to give the
relationships shown in Fig.~\ref{fig:plas}a between mutual information and
$\gamma_0^{CA1}$. The information is expressed in the figure as a fraction
of the information present when the pattern is stored in CA3 (\ref{eq:maxinf}).

Two phenomena can be seen in the results. The first, as mentioned in the
previous section (and discussed at more length in \citename{Tre95}, 1995),
is that information transmission is maximal when the plasticity of the
Schaffer collaterals is approximately matched with that of the preceding
stage of information processing. The second phenomenon is the increase in
information throughput with spread in the convergence about its mean. This
is an effect which is not immediately intuitive: it means that the
increase in mutual information provided by those CA1 neurons with a
greater number of connections than the mean more than compensates for the
decrease in those with less than the mean. It must be remembered that what
is being computed is the information provided by {\em all} CA1 cells about
patterns of activity in CA3. This increase in information is a network
effect that has no counterpart in the information a single CA1 cell could
convey. In any case, the effect is rather small: the realistic convergence
model allows the transmission of only marginally more information than the
uniform model. The uniform convergence approximation might be viewed as a
reasonable one for future analyses, then.

Fig.~\ref{fig:plas}b shows that the situation for graded pattern
distributions is almost identical. The numerical fraction of
information transmitted is of course lower (but total transmitted
information is similar -- see previous section). The uniform and
two-tier convergence models provide bounds between which the realistic
case must lie.

\section{Discussion and summary}

This chapter examined quantitatively the effect of analogue coding
resolution on the total amount of information that can be transmitted in a
model of the Schaffer collaterals. The tools used were analytical and
numerical, and the focus was upon relatively sparse codes. What can these
results tell us about the actual code used to signal information in the
mammalian hippocampus? In themselves, of course, they can make no definite
statement. It could be that there is a very clear maximum for information
transmission in using binary codes for the Schaffer collaterals, and yet
external constraints, such as CA1 efferent processing, might make it more
optimal overall to use analogue signalling. So results from a single
component study must be viewed with duep caution. However, these results
can provide a clear picture of the operating regime of the Schaffer
collaterals, and that is after all a major aim of any analytical study.

The results from this paper reiterate some previously known points,
and bring out others. For instance, it is very clear from
Fig.~\ref{fig:qlox} that, while nearly all of the information in the CA3
distribution can be transmitted using a binary code, this information
fraction drops off rapidly with analogue level. The total amount of
information transmitted is similar regardless of the amount of
analogue level to be signalled -- but this is a well known and
relatively general fact, and accords with common sense
intuition. However, the total amount of information that can be
transmitted is only {\em roughly} constant. It appears, from this
analysis, that while the total transmitted information drops off
slightly with analogue level for very sparse codes, the maximum moves
in the direction of more analogue levels for more evenly distributed
codes. This provides some impetus for making more precise measurements
of sparseness of coding in the hippocampus.

\begin{figure}
\epsfxsize=7cm
\epsfysize=6cm
\centerline{\epsffile{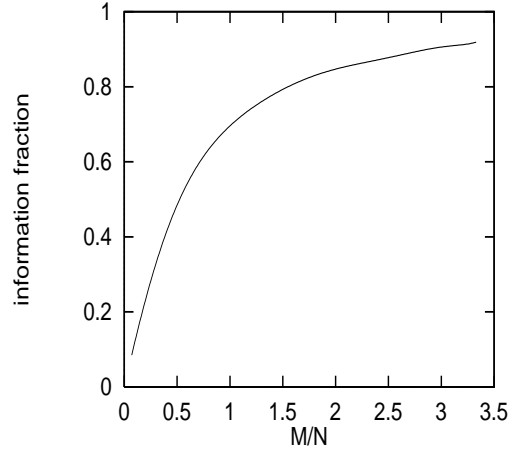}}
\caption{The dependence of information transmission on the expansion
ratio $r_{CA1,CA3} = M/N$.}
\label{fig:div}
\end{figure}

Another issue which this model allows us to address is the expansion ratio
of the Schaffer collaterals, i.e. the ratio between the numbers of neurons
in CA1 and CA3, $M/N$. It can be seen in Fig.~\ref{fig:div} that an
expansion ratio of 2 (a `typical' biological value) is sufficient for CA1
to capture most of the information of CA3, and that while the gains for
increasing this are diminishing, there is a rapid drop-off in information
transmission if it is reduced by any significant amount. The actual
expansion ratio for different mammalian species reported in the literature
is subject to some variation, with the method of nuclear cell counts
giving ratios between 1.4 (Long Evans rats) and 2.0 (humans) \cite{Ser88},
while stereological estimates range from 1.8 (Wistar rats) to 6.1 (humans)
\cite{Wes90}. It should be noted that in all these estimates, and
particularly with larger brains, there is considerable error (L. Seress,
personal communication). However, in all cases the Schaffer collateral
model appears to operate in a regime in which there is at least the scope
for efficient transfer of information.

Clearly it is essential to further constrain the model by fitting the
parameters as sufficient neurophysiological data becomes available. As
more parameters assume biologically measured values, the sensible ranges
of values that as-yet unmeasured parameters can take will become clearer.
It will then be possible to address further issues such as the
quantitative importance of the constraint upon dendritic length (i.e. the
number of synapses per neuron) upon information processing.

In summary, we have used techniques for the analysis of neural networks to
quantitatively investigate the effect of a number of biological issues on
information transmission by the Schaffer collaterals. We envisage that
these techniques, developed further and applied in a wider context to
networks in the medial temporal lobe, will yield considerable insight into
the organisation of the mammalian hippocampal formation.

\section*{Appendix A. Expression from the Replica Evaluation}

\begin{eqnarray}
\label{eq:inf}
\left<i\right> & = & {\rm extr}_{y_A,\tilde{y}_A} \Bigg\{ \sum_j \Gamma (y_A,w^0,z^0
,C_j,\gamma) - {N\over 2} y_A \tilde{y}_A \nonumber \\
&& + N\int D\tilde{s}_1 \left< F (\tilde{s}_1,0,\eta,\tilde{y}_A,0,0) 
\ln F (\tilde{s}_1,0,\eta,\tilde{y}_A, 0, 0) \right>_{\eta} \Bigg\} 
\nonumber\\
& - & {\rm extr}_{y_B,\tilde{y}_B,w_B,\tilde{w}_B,z_B,\tilde{z}_B}
\Bigg\{ \sum_j \Gamma (y_B,w_B,z_B,C_j,\gamma) \nonumber \\
&& -{N\over 2}(y_B\tilde{y}_B + 2 w_B \tilde{w}_B + z_B \tilde{z}_B)\nonumber\\
& &+N \int D\tilde{s}_1D\tilde{s}_2\left< F (\tilde{s}_1,\tilde{s}_2,
\eta,\tilde{y}_B,\tilde{w}_B,\tilde{z}_B ) \right>_{\eta} \nonumber\\
&& \times \ln \left<F(\tilde{s}_1,
\tilde{s}_2,\eta,\tilde{y}_B,\tilde{w}_B,\tilde{z}_B ) \right>_{\eta}\Bigg\} 
\end{eqnarray}

where taking the extremum means evaluating each of the two terms,
separately, at a saddle-point over the variables indicated. The notation
is as follows.  $N$ is the number of CA3 cells, whereas the sum over $j$
is over $M$ CA1 cells.  $F$ is given by

\eqa
F(\tilde{s}_1,\tilde{s}_2,\eta,\tilde{y},\tilde{w},\tilde{z})
 &=& \Bigg\{
\phi \left[{\eta +\sigma^2_{\delta}(\tilde{s}_+ -\tilde{w}\eta)\over 
\sigma_{\delta} \sqrt{ 1 + \sigma^2_{\delta} \tilde{y}} } \right]
{1\over \sqrt{ 1 + \sigma^2_{\delta} \tilde{y}}} \nonumber\\
&\times&\exp{\left[\eta +\sigma^2_{\delta}(\tilde{s}_+
-\tilde{w}\eta)\right]^2\over 
2\sigma^2_{\delta} (1 + \sigma^2_{\delta} \tilde{y} )}
+ \phi\left[{-\eta \over \sigma_{\delta} }\right]\exp {\eta^2\over 2
\sigma^2_{\delta}}\Bigg\} \nonumber\\
&\times&\exp \left[ \eta\tilde{s}_- -{\eta^2\over 2
\sigma^2_{\delta}}(1+\sigma^2_{\delta} \tilde{z} )\right]
\ena

and has to be averaged over $P_{\eta}$ and over the Gaussian variables of
zero mean and unit variance $\tilde{s}_1, \tilde{s}_2$.
\begin{equation}
Ds\equiv \frac{ds}{\sqrt{2\pi}}\exp -s^2/2\qquad \phi(x)\equiv
\int_{-\infty}^x Ds.\end{equation} $\tilde{y},\tilde{w}$ and $\tilde{z}$
are saddle-point parameters.  $\tilde{s}_+$ and $\tilde{s}_-$ are linear
combinations of $\tilde{s}_1,\tilde{s}_2$:
\begin{equation}
\tilde{s}_{\pm}=\sum_{k=1}^2 (\mp 1)^{(k-1)}\sqrt{\left[\sqrt{(\tilde{y}-
\tilde{z})^2+4\tilde{w}^2}\mp (-1)^k(\tilde{y}-\tilde{z})\right](\tilde{y}
\tilde{z}-\tilde{w}^2) \over \left[ \tilde{y}+\tilde{z}+(-1)^k\sqrt{(\tilde{y}
-\tilde{z})^2+4\tilde{w}^2}\right]\sqrt{(\tilde{y}-\tilde{z})^2+
4\tilde{w}^2} }\tilde{s}_k
\end{equation}
in the last two lines of Eq.~\ref{eq:inf}, but in the second line of
Eq.~\ref{eq:inf} one has $\tilde{s}_+ = \tilde{s}_1\sqrt{\tilde{y}_A},
\tilde{s}_- = 0$.

$\Gamma $ is effectively an entropy term for the CA1 activity distribution,
given by
\begin{eqnarray}
\Gamma (y,w,z,C_j,\gamma)&=& \int {ds_1ds_2\over 2\pi \sqrt{\det 
{\bf T}_j'}}\exp - (\begin{array}{cc}s_1& s_2 \end{array}) {({\bf T}_j')^{-1}
\over 2}\left(\begin{array}{c}s1\\s2\end{array}\right)\nonumber\\
&&\times\left[\int_{-\infty}^0
dUG(U)\ln\int_{-\infty}^0dU'G(U') \right.\nonumber\\
&&+\left.\int^{\infty}_0dUG(U)\ln G(U)\right],
\end{eqnarray}
where
\begin{eqnarray}
G(U)&=&G(U;s_1,s_2,y,w,z,C_j,\gamma)\nonumber\\
&=&\phi\left[{(\xi_0-s_2)
(T_{yj}+2g_jT_{wj}+g^2_jT_{zj})+(U-U_0+s_1+g_j
s_2)(T_{wj}+g_jT_{zj})\over \sqrt{(T_{yj}T_{z_j}-T^2_{wj})
(T_{yj}+2g_jT_{wj}+g^2_jT_{zj})}}\right]\nonumber\\
&&\times {1\over\sqrt{2\pi(T_{yj}+2g_jT_{wj}+g^2_jT_{zj})}}
\exp -{(U-U_0+s_1+g_js_2)^2\over 2(T_{yj}+2g_jT_{wj}+g^2_j
T_{zj})}\nonumber\\
&+&\phi\left[{-(\xi_0-s_2)T_{yj}-(U-U_0+s_1+g_j\xi_0)T_{wj}\over 
\sqrt{(T_{yj}T_{z_j}-T^2_{wj})T_{yj}}}\right]\nonumber\\
&&\times {1\over\sqrt{2\pi T_{yj}}}
\exp -{(U-U_0+s_1+g_j\xi_0)^2\over 2T_{yj}},
\end{eqnarray}
and
\begin{eqnarray}
T_{yj}&=&\sigma^2_{\epsilon^R}+\sigma^2_JC_j(y^0-y)\nonumber\\
T_{wj}&=&\sigma^2_JC_j(w^0-w)\cos (\theta)\nonumber\\
T_{zj}&=&\sigma^2_{\epsilon^S}+\sigma^2_JC_j(z^0-z)\\
{\bf T}_j'&=&\sigma^2_JC_j\left(\begin{array}{cc}y&w\cos (\theta)\\
w\cos (\theta)&z\end{array}\right)\nonumber
\end{eqnarray}
are effective noise terms.
\begin{equation}g_j =  h {C_j\over C}x^0\left<\eta \right>_{\eta}\sqrt{C\gamma (\theta)},
\end{equation} 
$y,w,z$ are saddle-point
parameters (conjugated to $\tilde{y},\tilde{w}$ and $\tilde{z}$), and
$x^0,y^0,w^0,z^0$ are corresponding single-replica parameters fixed as
\begin{eqnarray}
x^0&=&{1\over N}\sum_i \left< {(\eta_i -\left<\eta \right>_{\eta})\over \left<\eta \right>_{\eta
}}V_i \right> \nonumber\\ 
&=& \left< {(\eta -\left<\eta \right>_{\eta})\over \left<\eta \right>_{\eta }}\left[\eta\phi\left(
{\eta\over\sigma_{\delta}}\right)+{\sigma_{\delta}\over\sqrt{2\pi}}\exp
-{1\over 2}\left({\eta\over\sigma_{\delta}}\right)^2\right]\right>_{\eta}
\nonumber\\
y^0&=&{1\over N}\sum_i \left< V_i^2\right> 
=\left<\left[\sigma_{\delta}^2+\eta^2\right]\phi
\left({\eta\over\sigma_{\delta}}\right)+{\eta\sigma_{\delta}\over\sqrt{2\pi}}
\exp-{1\over 2}\left({\eta\over\sigma_{\delta}}\right)^2\right>_{\eta}
\nonumber\\
w^0&=&{1\over N}\sum_i \left< \eta_i V_i\right> = \left< \eta\left[\eta\phi\left(
{\eta\over\sigma_{\delta}}\right)+{\sigma_{\delta}\over\sqrt{2\pi}}
\exp-{1\over
2}\left({\eta\over\sigma_{\delta}}\right)^2\right]\right>_{\eta}
\nonumber\\
z^0&=&{1\over N}\sum_i \eta_i^2 = \left< \eta^2 \right>_{\eta}.
\end{eqnarray}

\section{Appendix B. Parameter Values}

Parameters used were, except where otherwise indicated in the text:

\begin{center}
\begin{tabular}{|l|l|}
\hline
$\sigma_\delta$ & 0.30 \\ \hline
$\sigma_\epsilon^S$ & 0.20 \\ \hline
$\sigma_\epsilon^R$ & 0.20 \\ \hline
$ C $ & 10000 \\ \hline
$\sigma_J^2$ & 0.0001 \\ \hline
$\xi_0$ & -0.4 \\ \hline
$U_0$ & -0.4 \\ \hline
$M/N$ & 2.0 \\ \hline
\end{tabular}
\end{center}

\bibliography{srs}

\end{document}